\documentclass[preprint,pre, floats,aps,amsmath,amssymb]{revtex4}
\usepackage{bm}
\usepackage[colorlinks=true,allcolors=blue]{hyperref}
\usepackage{lipsum} 
\usepackage{graphicx}
\usepackage{subfig}
\usepackage{cleveref}
\usepackage{mathtools} 
\usepackage{soul} 
\usepackage{float}
\usepackage[normalem]{ulem}
\usepackage{multirow}
\usepackage{amsfonts}
\newcommand{\overbar}[1]{\mkern 1.5mu\overline{\mkern-1.5mu#1\mkern-1.5mu}\mkern 1.5mu}

\begin{document}                  
\title{Self-consistent gyrokinetic modelling of turbulent and neoclassical tungsten transport in toroidally rotating plasmas}
\author{K. Lim$^1$, X. Garbet$^{2,4}$, Y. Sarazin$^2$, E. Gravier$^3$, M. Lesur$^3$, G. Lo-Cascio$^3$ and T. Rouyer$^3$}
\affiliation{$^1$École Polytechnique Fédérale de Lausanne (EPFL), Swiss Plasma Center (SPC), CH-1015 Lausanne, Switzerland \\
$^2$CEA, IRFM, F-13108 Saint-Paul-L\`ez-Durance, France \\
$^3$Institut Jean Lamour (IJL), UMR 7198 CNRS - Université de Lorraine, 54000 Nancy, France \\
$^4$School of Physical and Mathematical Sciences, Nanyang Technological University, 637371 Singapore}

\begin{abstract}
The effect of toroidal rotation on both turbulent and neoclassical transport of tungsten (W) in tokamaks is investigated using the flux-driven, global, nonlinear 5D gyrokinetic code GYSELA. Nonlinear simulations are carried out with different levels of momentum injection that drive W to the supersonic regime, while the toroidal velocity of the main ions remains in the subsonic regime. The numerical simulations demonstrate that toroidal rotation induces centrifugal forces that cause W to accumulate in the outboard region, generating an in-out poloidal asymmetry. This asymmetry enhances neoclassical inward convection, which can lead to central accumulation of W in cases of strong plasma rotation. The core accumulation of W is mainly driven by inward neoclassical convection. However, as momentum injection continues, roto-diffusion, proportional to the radial gradient of the toroidal velocity, becomes significant and generate outward turbulent flux in the case of ion temperature gradient (ITG) turbulence. Overall, the numerical results from nonlinear GYSELA simulations are in qualitative agreement with the theoretical predictions for impurity transport, as well as experimental observations.
\end{abstract}
\maketitle
\section{Introduction}
Tungsten (W), a high-Z element, is considered for use in the first-wall plasma-facing components in future tokamaks due to its high melting point, low erosion rate, and good thermal conductivity \cite{Kaufmann2007}. However, experiments have shown that the accumulation of W in the core region is frequently observed due to neoclassical inward convection, leading to a reduced plasma performance and energy loss through radiation \cite{Tokar1997, Kochl2018}. In particular, radiative loss can cause complete extinction of the burning plasma, even at low W concentration \cite{Putterich2010}.  Consequently, it is crucial to mitigate the accumulation of W in tokamaks to improve efficiency and performance of the reactor.

Over the past few decades, a strong emphasis has been placed on the non-uniform distribution of impurity density over magnetic flux surfaces \citep{Wesson1997, Romanelli1998, Bilato2014, Angioni2014, Maget2020}. While conventional neoclassical theory often assumes a uniform density distribution \citep{Hinton1976, Hirshman1981}, it is now widely accepted that an accurate prediction of neoclassical impurity transport requires a proper estimation of the poloidal asymmetry of impurity density as it can enhance or reduce neoclassical impurity transport \cite{Casson2010, Casson2013, Angioni2014, Belli2014}.

The toroidal rotation of plasma in tokamaks, whether driven intrinsically \citep{Bortolon2006, Rice2007} or externally by neutral beam injection (NBI) heating, often results in a non-uniform distribution of W density. In particular, the centrifugal force generated by toroidal rotation drives impurities towards the outer wall, yielding a strong 'in-out' poloidal asymmetry that leads to the enhancement of neoclassical transport \citep{Helander1998, Angioni2014, Casson2015, Breton2018}. Although the impact of toroidal rotation on turbulent transport is found to be much smaller than on neoclassical transport \citep{Casson2015}, the turbulent roto-diffusion generated by a radial gradient in the toroidal rotation \cite{Camenen2009} and the parallel velocity gradient (PVG) instabilities \cite{Catto1973} can play an important role in the case of a strongly rotating plasma.

The impact of toroidal rotation on both turbulent and neoclassical W transport was investigated in Ref. \cite{Casson2015}. The main approach used in the study was to compute turbulent and neoclassical transport separately. Turbulent transport was calculated using the gyrokinetic model in the toroidally co-moving frame, while neoclassical transport was evaluated using the drift-kinetic code NEO, but without considering frictional effects at the lowest order \cite{Hinton1985, Belli2014}.  

The aim of this paper, as an extension of the previous work on impurity transport from light to heavy impurities \cite{Lim2021}, is to explore the effect of toroidal rotation on turbulent and neoclassical W transport using the full-F gyrokinetic code GYSELA \cite{Grandgirard2016}. By employing the multi-species collisional operator implemented in GYSELA \cite{Donnel2019}, we investigate the self-consistent treatment of both neoclassical and turbulent transport channels. The interplay between these channels can have a significant impact on impurity transport and its accumulation in the core of tokamaks \cite{Esteve2018, Idomura2021}. To this end, we examine the impact of sonic (for the impurities) toroidal rotation on turbulent and neoclassical transport in the presence of poloidal asymmetry enhanced by centrifugal forces. 

The remainder of the paper is organized as follows. Firstly, in Section \ref{Sec2}, we introduce the GYSELA code and the source terms for studying impurity transport in a rotating plasma. Section \ref{Sec3} discusses the results of the nonlinear GYSELA simulations, with a focus on the poloidal asymmetry generated by the external momentum injection. In Section \ref{Sec4}, the effect of sonic rotation on both turbulent and neoclassical transport is explored, highlighting the inward neoclassical convection due to the poloidal asymmetry and the outward roto-diffusion convection due to the steepening of the velocity gradient. Finally, the paper concludes with a discussion of the main findings and conclusions in Section \ref{Sec5}.

\section{Numerical model}\label{Sec2}
The GYSELA code is a first-principles, nonlinear, global and flux-driven 5D gyrokinetic code that evolves the gyro-averaged distribution function, $\bar{F}_s$, for each species $s$, in time and phase space $\mathbf{z}=(\mathbf{x}_G, v_{G\parallel}, \mu_G)$, $\mathbf{x}_G$ being the gyro-center position, $v_{G\parallel}$ the parallel velocity and $\mu$ the magnetic moment, based on a semi-Lagrangian method for numerical computations \cite{Grandgirard2016, Garbet2010}. The nonlinear gyrokinetic Vlasov equation is expressed as
\begin{align}
    \frac{\partial \bar{F}_s}{\partial t} + \frac{1}{B_{\parallel}^*}\nabla_{\textbf{z}} \cdot (\dot{\textbf{z}}B_{\parallel}^* \bar{F_s})=\mathcal{C}(\bar{F}_s) + \mathcal{S}_{\text{heat}}+\mathcal{S}_{\text{\text{mom}}},
\label{Sec2:eq_gysela}
\end{align}
where we use $B_\parallel^*=\mathbf{b}
\cdot \mathbf{B}_\parallel^*$ with the generalized magnetic field $\mathbf{B}_\parallel^*=\mathbf{B}+(m_s  v_{G\parallel}/eZ_s) \nabla \times \mathbf{b}$, the charge number and mass,  $m_s, Z_s$, the collision operator $\mathcal{C}(\bar{F_s})$ and $\mathcal{S}_{\text{heat}}, \mathcal{S}_{\text{\text{mom}}}$ represent the heat and toroidal momentum sources respectively. The evolution of $\bar{F}_s$ is self-consistently coupled to the quasi-neutrality equation to ensure the balance between the total density $N_{eq}$ and the electric potential $\phi$. In the limit of long wavelengths and assuming adiabatic electrons, the electrostatic quasi-neutrality equation can be written as
\begin{align}
    e\Bigg(\frac{\phi-\langle \phi \rangle}{T_e} \Bigg)-\sum_s\frac{1}{N_{eq,s}} \nabla_\perp \cdot \Bigg( \frac{m_sN_{eq,s}}{eZ_sB^2 }\nabla_\perp \phi\Bigg) &= \sum_s \frac{1}{N_{eq,s}}\int d\mathbf{v} \mathcal{J}(\bar{F}_s-\bar{F}_{eq,s}),
    \label{Sec2:quasi_neutrality}
\end{align}
where $\langle \dots \rangle$ denotes the flux surface average, $T_e$ is the electron temperature, $\int d\mathbf{v}=\mathcal{J}_v d\mu dv_{G\parallel}$ with $\mathcal{J}_v=(2\pi B_\parallel^*/m_s)$ and $\mathcal{J}$ is the gyro-average operator. In the above expressions, the mass $m$, the magnetic field $B$, the density $N$, the temperature $T$ and the electric potential $\phi$ are normalized with respect to references values $m_0, B_0, N_0, T_0$ and $T_0/eZ_0$. The length scales are normalized to the reference Larmor radius $\rho_0=v_{T_0}/\omega_{c0}$, where $v_{T_0}=\sqrt{T_0/m_0}$ and $\omega_{c0}=eZ_0B_0/m_0$, and the time is normalized to $\omega_{c0}^{-1}$. All drift velocities are normalized to $v_{T_0}$, while the toroidal velocity is normalized to $v_{T_0,s}=v_{T_0}\sqrt{m_0/m_s}$. Moreover, for the large aspect ratio limit, we assume that the parallel velocity of the particles in GYSELA is equivalent to the toroidal velocity. Under this assumption, the parallel velocity measured in GYSELA can be interpreted as the Mach number, which we define as $M_s=v_{\varphi,s}/v_{th,s}$.

The presence of the heat source $\mathcal{S}_{\text{heat}}$ in Eq. (\ref{Sec2:eq_gysela}) is essential to allow flux-driven simulations to reach a statistical steady state. External torque is injected to the plasma via the momentum source $\mathcal{S}_{\text{mom}}$. Each of these sources is designed so as to have a non-vanishing contribution to a single fluid moment equation, related to heat and parallel momentum in the present study.

In this paper, as derived in Ref. \citep{Sarazin2011}, we adopt the following form of heat and momentum sources
\begin{align}
        \mathcal{S}_{\text{heat}} &=\frac{S_E S_r}{\sqrt{2}\pi^{3/2}T_{sce}^{5/2}}\Big[\bar{v}_{G\parallel}^2-\frac{1}{2}-\frac{J_{\parallel B}}{2-J_{\parallel B}^2}(2-\bar{\mu})(2\bar{v}_{G\parallel}-J_{\parallel B})\Big]e^{-\bar{v}_{G \parallel}^2 -\bar{\mu}},
\label{Sec2:heat_source}
\end{align}
where we use the normalized parallel velocity $\bar{v}_{G\parallel} = v_{G\parallel}/\sqrt{2T_{sce}/m_s}$, the normalized magnetic moment $\bar{\mu}=\mu B / T_{sce}$, $J_{\parallel B}=\sqrt{2T_{sce}}J_\parallel / B^2$ with the normalized source temperature $T_{sce}$, the parallel current $J_\parallel$, the radial envelope profile $S_r$ and the amplitude of the source $S_E$. Similarly to the heat source, the analytical expression of the momentum source can be recast as
\begin{align}
    \mathcal{S}_{\text{mom}} &= \frac{S_{mom}S_r}{4\pi^{3/2}T_{sce}^2}[2\bar{v}_{G\parallel}(2-\bar{\mu})-J_{\parallel B}(1+2\bar{v}_{G\parallel}^2-\bar{\mu})]e^{-\bar{v}_{G \parallel}^2 -\bar{\mu}},
\label{Sec2:mom_source}
\end{align}
where $S_{mom}$ is the amplitude of the momentum source that acts as a control parameter of the simulations to achieve different toroidal rotations in the nonlinear simulations. The radial profile of $S_r$ is composed of two hyperbolic tangents
\begin{align}
    S_r(\rho) = -\frac{1}{2}\Bigg[\tanh\Bigg( \frac{\rho-\rho_s-3L_s}{L_s}\Bigg) + \tanh\Bigg( \frac{\rho_s-3L_s-\rho}{L_s}\Bigg)  \Bigg],
\label{Sec2:source_radial_profile}
\end{align}
where we use $\rho=r/a$, the radial position of the source $\rho_s$ and the width of the source $L_s$. Although off-axis injection of NBI is known to provide control over density and current profiles \cite{Dux2003}, the present study concentrates on central injection of toroidal momentum to achieve consistent gradient profiles with experimental observations.

In a toroidally rotating frame with the velocity $\mathbf{u}$, the particle velocity $\mathbf{v}$ in the laboratory reference frame is expressed as $\mathbf{v}=\Omega R^2 \nabla \varphi + \mathbf{u}$ with the toroidal rotation frequency $\Omega$, and the Hamiltonian is defined as follows
\begin{align}
    H = \frac{1}{2}mu_\parallel^2 + \mu B + eZ\phi - \frac{m \Omega^2 R^2}{2},
\label{Sec2:Hamiltonian}
\end{align}
where the electrostatic potential is decomposed into, $\phi=\phi_0(\psi) + \tilde{\phi}_1(\psi, \theta)$, an equipotential part on the flux surface $\phi_0(\psi)$, and a poloidally varying part $\tilde{\phi}_1(\psi, \theta)$ introduced to ensure the quasi-neutrality in the presence of centrifugal forces. 

According to the neoclassical theory, developed in Refs. \citep{Hinton1985, Wong1987}, at the lowest order in $\delta=\rho_{i}/L$, where $\rho_{i}$ represents the ion Larmor radius and $L$ denotes the radial length scale, the distribution function is a Maxwellian. In addition, the temperature $T_{eq}(\psi)$ and toroidal rotation frequency $\Omega(\psi)$ are flux functions, which are constant on given magnetic flux surfaces, while the density $N_{eq}(\psi,\theta)$ is not a flux function due to the centrifugal force. Consequently, the Maxwellian function in the rotating frame can be expressed as
\begin{align}
    F_{M} = N(\psi,\theta)\bigg(\frac{m}{2\pi T_{eq}(\psi)}\bigg)^{3/2} \exp{\Bigg(-\frac{m u_\parallel^2}{2T_{eq}(\psi)} -\frac{\mu B_{eq}(\psi,\theta)}{T_{eq}(\psi)}\Bigg)}.
\end{align}
Using the Hamiltonian derived in Eq. (\ref{Sec2:Hamiltonian}), the above expression can be rewritten as 
\begin{align}
    F_{M} &= N_{eq}(\psi)\bigg(\frac{m}{2\pi T_{eq}(\psi)}\bigg)^{3/2} \exp{\bigg(-\frac{H}{T_{eq}(\psi)}\bigg)}, 
\label{Sec2:Maxwellian}
\end{align}
with the expression of the density distribution in a toroidally rotating plasma
\begin{align}
    N(\psi, \theta)=N_{eq}(\psi)\exp{\Bigg(-\frac{eZ\phi_0(\psi)}{T_{eq}(\psi)}-\frac{eZ\tilde{\phi}_1(\psi,\theta)}{T_{eq}(\psi)}+\frac{1}{2}\frac{m\Omega^2R^2}{T_{eq}(\psi)}\Bigg)}.
\label{Sec2:Density_distribution}
\end{align}
In the presence of strong rotation, impurities tend to accumulate in the outboard region due to the centrifugal force, while an induced equilibrium potential $\tilde{\phi}_1$, which varies poloidally, acts to balance the resulting density asymmetry. These effects are more pronounced for heavy impurity ions because of their large mass and high-Z.

It should be noted that, unlike gyrokinetic formulation in the co-moving frame that requires the inclusion of both the Coriolis drift and centrifugal effect in the derivation of guiding-center equations \cite{peeters2009, Brizard1995, Sugama2017}, the equations of motion in GYSELA are solved in the laboratory frame. This implies that the effect of toroidal rotation is inherently present in the distribution function described in Eq. (\ref{Sec2:Maxwellian}), without introducing any additional terms in the dynamics of the guiding-center.
\section{Numerical results}\label{Sec3}
In this section, we describe a set of nonlinear GYSELA simulations to explore the effect of toroidal rotation on impurity transport. To this end, we consider W as the main impurity due to its large mass $A_W=184$ and high charge state $Z_W=74$ when fully ionized. However, we restrict our analysis to the partially ionized state $Z_W=40$ as W is not fully stripped even at the core temperature of tokamaks. We conduct the simulations in a circular geometry with a numerical resolution of $N_r \times N_\theta \times N_\varphi \times N_{v_{\parallel}} \times N_\mu=256 \times 512 \times 32 \times 128 \times 64$, using a time step of $\Delta t = 16 \omega_{ci}^{-1}$. The nonlinear gyrokinetic Vlasov equation defined in Eq. (\ref{Sec2:eq_gysela}) is solved for both main ions and W, while electrons are simplified to follow a Boltzmann response within a flux surface. The initial radial profiles of density and temperature are established at $R/L_{n}=2.2$ and $R/L_{T}=6$ for both species, triggering ITG turbulence. Other parameters such as normalized gyro-radius $\rho_* = \rho_i/a=1/190$, inverse aspect ratio $\epsilon=a/R_0\simeq 0.2$, safety factor $q_{95}\simeq 3$ are kept constant throughout the present study. The radial profile of collisionality is determined by the normalized collisionality $\nu^*_s = \nu_i q R_0/(v_{T_s} \epsilon^{3/2})$. For the analysis, we set the main ions in the banana regime with $\nu_D^*=0.1$ and W in the Pfirsch-Schlüter (PS) regime with $\nu_W^*\simeq 23$ near the edge. The W concentration is restricted to the trace limit ($C_W = 10^{-6}$), ensuring that the background turbulence remains unaffected by its presence \cite{Lesur2020}. 

\begin{figure}[H]
\begin{center}    
\includegraphics[width=0.4\textwidth]{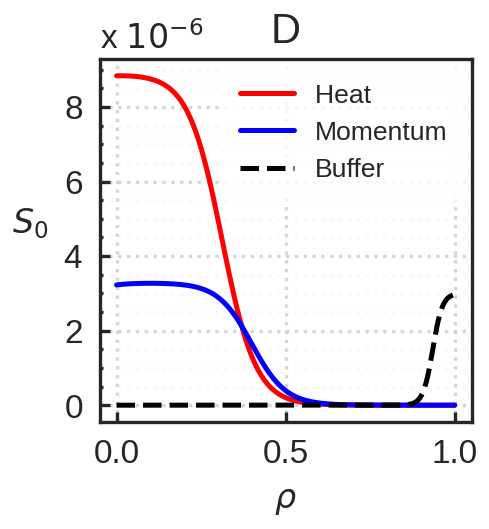}
\caption{Radial profiles of heat and momentum sources applied to the main ions in GYSELA. No heat and momentum sources are applied to W. The buffer region is implemented to prevent boundary effects from affecting the bulk plasma.}	
\label{Fig:radial_source_profile}
\end{center}
\end{figure}

In this study, increasing values of $S_{mom}$ allow one to reach increasing values of Mach number. The radial profile of the momentum source remains unchanged with $\rho_{s,\text{mom}}=0.1$ and $L_{s,\text{mom}}=0.1$. In addition, the amplitude and radial profile of the heat flux $\mathcal{S}_{\text{heat}}$ remain constant throughout the analysis, with $S_E=0.017$, $\rho_{s,\text{heat}}=0.012$ and $L_{s,\text{heat}}=0.1$. Heat and momentum are injected only for deuterium, indicating that the transfer of heat and kinetic energy occurs via inter-species collisional effects. Figure \ref{Fig:radial_source_profile} shows the radial profile of the different source terms for the case with ${S}_{mom}=10^{-2}$. To mitigate the impact of steep gradients at the boundary of the simulation domain, a numerical buffer region with a diffusive term is introduced near the edge.

To investigate its effect on density distribution and impurity transport that is typically proportional to $\propto (1+M^2)$ \cite{Romanelli1998}, W ions are accelerated to supersonic regime. Although the Mach number of main ions in fusion plasma is low, typically $M_D < 0.4$, the Mach number of W can exceed the unity $(M_W > 1)$ due to the mass ratio $M_W = \sqrt{A_W/A_D}M_D$.

The numerical computation time required to reach a steady state is proportional to the gyro-Bohm scaling $\propto \rho_*^{-3}$ \cite{Waltz1990}, which makes it infeasible to attain a steady state for multi-species simulations. To overcome this limitation, we first perform a simulation without W impurities for a sufficiently long time $(\sim 10^5 \omega_{ci}^{-1})$ to develop the background turbulence and saturate the toroidal velocity of main ions. Then, we carry out the second part of the simulation with W impurities for more than $10^5\omega_{ci}^{-1}$ to get convergence on the toroidal rotation of W through collisional effects. Once the simulations are close to the quasi-steady state, all quantities are analyzed with a time window of $\sim 3000\omega_{ci}^{-1}$ and are averaged over flux surfaces. 
\begin{figure}[H]
\begin{center}
    \includegraphics[width=\textwidth]{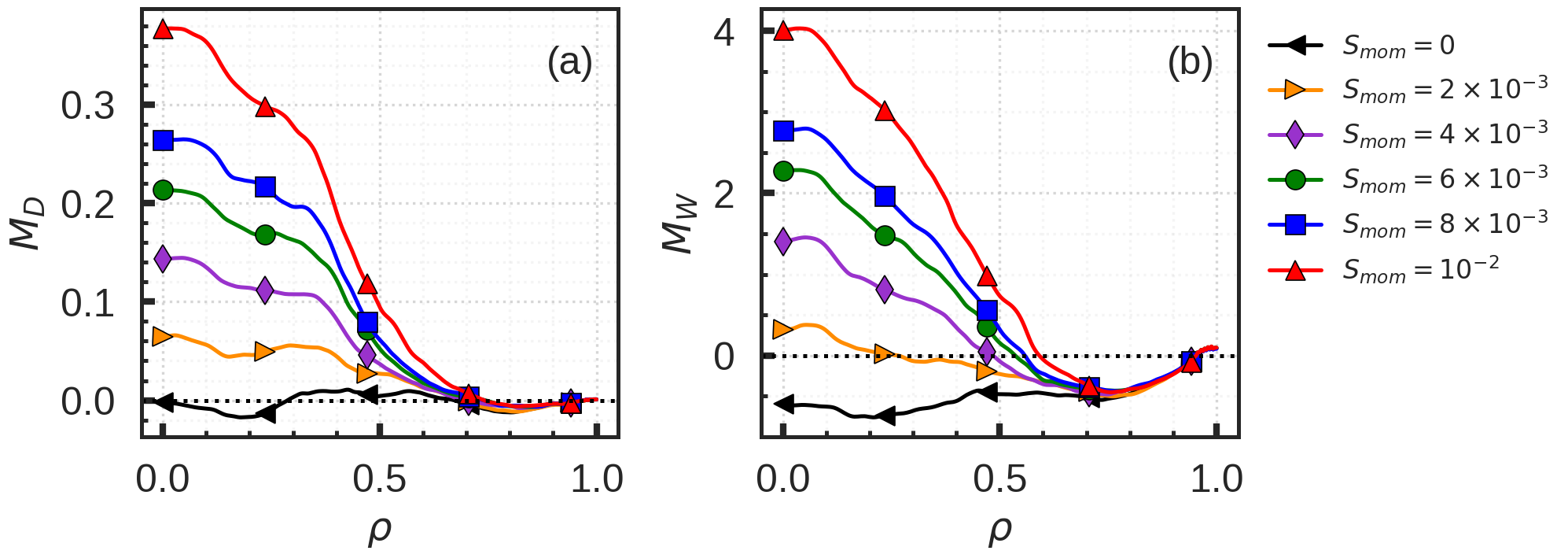}
\caption{Mach numbers for the main ions (a) and W (b) with different amplitudes of momentum sources plotted as a function of $\rho=r/a$ at $t=2\times 10^5 \omega_{ci}^{-1}$.}
\label{Fig:Radial_Vtor}
\end{center}
\end{figure}

The radial profiles of the Mach number for both species are illustrated in Figure \ref{Fig:Radial_Vtor} with different amplitudes of the momentum source. In the absence of the momentum injection $S_{mom}=0$, the toroidal velocity, produced by turbulent intrinsic rotation \cite{Bortolon2006, Rice2007}, exhibits low and flat profile for both species in the counter-current direction, i.e. $M_D \sim -0.02$ and $M_W \sim -0.6$. Once the momentum source is activated, the toroidal rotation increases and converges to a similar shape as the source profile illustrated in Figure \ref{Fig:radial_source_profile}. W is propelled into the supersonic regime for amplitudes stronger than $S_{mom}=4\times 10^{-3}$. In the core region, the value of $M_W$ reaches up to $M_W \simeq 4$ with $S_{mom}=10^{-2}$. Despite the injection of momentum, the Mach number and its gradient for deuterium remains at low levels, with values of $M_D < 0.4$ and $\lvert dv_{\varphi,D}/dr \rvert \simeq 0.4$, respectively. According to the criterion for the PVG instability to occur, which is $\lvert dv_{\varphi,D}/dr \rvert > R/(L_n\sqrt{\tau+1})$ with $\tau=T_{e}/T_i$ \cite{Catto1973}, the present nonlinear simulations are not subject to PVG instabilities. Similar Mach numbers have been observed in JET experiments with NBI injection \cite{vries2008}, where the Mach number for deuterium was found to be as high as $M_D \simeq 0.7$, yielding $M_W \simeq \sqrt{A_W/A_D} M_D \simeq 4.5$.

\begin{figure}[H]
\begin{center}
\includegraphics[width=\textwidth]{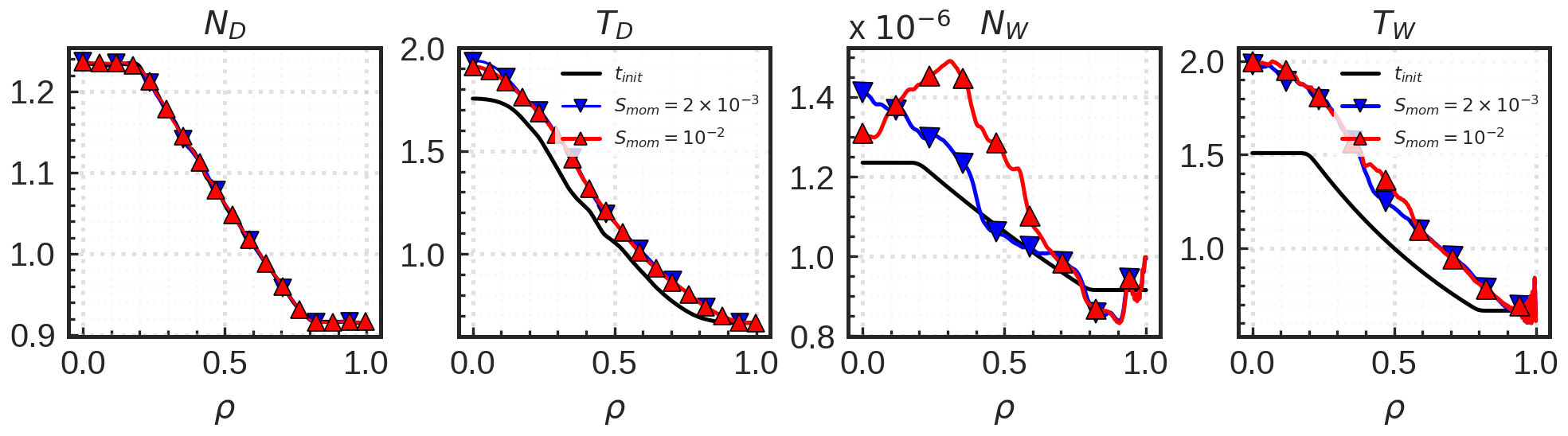}
\caption{Radial density and temperature profiles of main ions and W with different values of momentum source measured at $t \simeq 2\times 10^{5}\omega_{ci}^{-1}$. The black solid line represents the initial profile at $t\simeq10^{5}\omega_{ci}^{-1}$ where W is introduced.}
\label{Fig:Radial_NTP_profiles}
\end{center}
\end{figure}

The radial profile of the flux-averaged density and temperature for both the main ions and W are displayed in Figure \ref{Fig:Radial_NTP_profiles}. For deuterium, the density and temperature profiles are not significantly affected by the external momentum injection with the considered amplitudes of $2\times 10^{-3} \leq S_{mom} \leq 10^{-2}$. The slight increase in $T_D$ is likely due to the balance between the heat source and the outward heat flux, rather than the effect of toroidal rotation. This indicates that the injected momentum does not have a stabilizing effect on the turbulence, and the intensity of the background ITG turbulence remains consistent throughout this study. Previous works with GYSELA have demonstrated the stabilizing effect on turbulence via the formation of transport barrier generated by a vorticity source term \cite{Strugarek2011, Lo-Cascio2022}. Furthermore, recent research with the JET-ILW (ITER-like wall) tokamak reports that steep temperature gradients at the plasma periphery can prevent W accumulation in the core region by enhancing outward neoclassical convection \cite{Field2023}. The impact of transport barriers on impurity transport using GYSELA is currently under investigation.

Unlike deuterium, we observe significant changes in W density as we increase the magnitude of $S_{mom}$, while the temperature profiles remain unaffected and saturate at the same level as the main ions, indicating energy transfer by collisional effects. The overall density profile suggests a central accumulation in cases of strong rotation. More detailed evidence of these effects is presented in the following sections.

\subsection*{Poloidal asymmetry of W density in toroidally rotating plasma}
\begin{figure}[H]
\begin{center}
\includegraphics[width=1\textwidth]{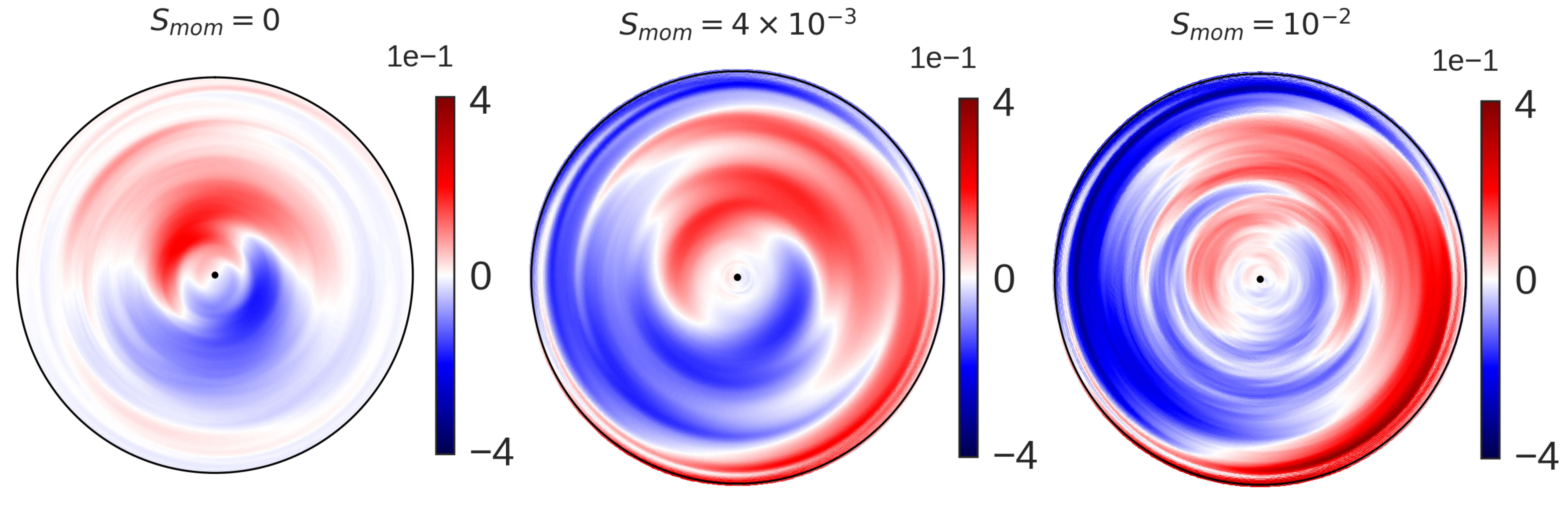}
\caption{2D poloidal section of perturbed W density obtained from the nonlinear GYSELA simulations. Three different cases are represented, i.e. without momentum source $S_{mom}=0$ (left), the Mach number close to unity $M_W \sim 1$ with $S_{mom}=4 \times 10^{-3}$ (center), and the high Mach number $M_W \simeq 4$ with $S_{mom}=10^{-2}$ (right).  }
\label{Fig:Asymmetry_GYSELA}
\end{center}
\end{figure}

Non-uniform distributions of W obtained from the GYSELA simulations of rotating plasma are shown in Figure \ref{Fig:Asymmetry_GYSELA}. Three different cases are compared to elucidate the effect of toroidal rotation in driving an enhanced in-out asymmetry. In the absence of momentum injection $S_{mom}=0$, an up-down asymmetry of W is observed mainly due to the background turbulence \citep{Donnel2019_b}. The degree of asymmetry is approximately 20\%. As the amplitude of the source term increases, the geometry of asymmetry changes from up-down to in-out due to centrifugal effects. In particular, the case with $S_{mom}=10^{-2}$ yields an excess of approximately 40\% of W being localized in the outboard region.

An analytical formula for non-uniform density distribution in a toroidally rotating plasma can be obtained by solving the drift-kinetic equation for impurities in the Pfirsch-Schlüter regime \cite{Fulop1999}. By assuming an axisymmetric system and small inverse aspect ratio $(\epsilon \ll 1)$, the poloidal asymmetry of impurity density at the lowest order of $\epsilon$ can be written as
\begin{align}
    N_W=1+N_c \cos\theta + N_s \sin\theta,
\label{Eq:Fulop}
\end{align}
with the up-down geometrical factor
\begin{align}
    N_s = 2\epsilon g \frac{1+(1+\gamma)M_0^2}{1+(1+\gamma)^2g^2}
    \label{Eq:Fulop_ud}
\end{align}
and the in-out geometrical factor
\begin{align}
     N_c = 2\epsilon \frac{M_0^2-(1+\gamma)g^2}{1+(1+\gamma)^2g^2}.
     \label{Eq:Fulop_io}
\end{align}
In the above expressions, the following dimensionless parameters are introduced, such as $M_0^2 = M^2 R_0^2 / R^2$ with
\begin{align}
    M^2 = \frac{m_z \omega^2 R^2}{2T_z}\bigg(1-\frac{Zm_i}{m_z} \frac{T_e}{T_e+T_i}\bigg)
\label{Eq:modified_mach_number}
\end{align}
and
\begin{align}
    g = -\frac{m_i Z^2 I}{e\tau_{iz}\langle \bm{B}\cdot \nabla \theta \rangle} \bigg( \frac{1}{N_i}\frac{\partial N_i}{\partial \psi} -\frac{1}{2}\frac{1}{T_i}\frac{\partial T_i}{\partial \psi}\bigg),
\end{align}
where we use
\begin{align}
    \gamma &= \frac{eL_\perp \langle B^2 \rangle u}{T_i}, \\
    L_\perp^{-1} &=-I\bigg(\frac{1}{N_i}\frac{\partial N_i}{\partial \psi}- \frac{3}{2}\frac{1}{T_i}\frac{\partial T_i}{\partial \psi}\bigg), 
\end{align}
and the poloidal rotation for the main ions
\begin{align}
    u\sim -0.33\frac{I f_c} {e\langle B^2 \rangle}\frac{\partial T_i}{\partial \psi}.
\end{align}
with the fraction of circulating particles $f_c \simeq 1-1.46\sqrt{\epsilon}$. 

The first dimensionless parameter, defined in Eq. (\ref{Eq:modified_mach_number}), corresponds to a modified Mach number. In Eqs. (\ref{Eq:Fulop_ud}, \ref{Eq:Fulop_io}), the geometrical factors scale quadratically with respect to $M$. The second dimensionless parameter, denoted as $g$, characterizes the steepness of the density and temperature gradient profiles for the main ions. In standard neoclassical theory, this parameter is assumed to be small, indicating a weak frictional force. However, its impact becomes dominant when impurities localize in the outboard region, particularly for high-Z impurities ($g \propto Z^2$) or in the pedestal region.

\begin{figure}[H]
\begin{center}
\includegraphics[width=\textwidth]{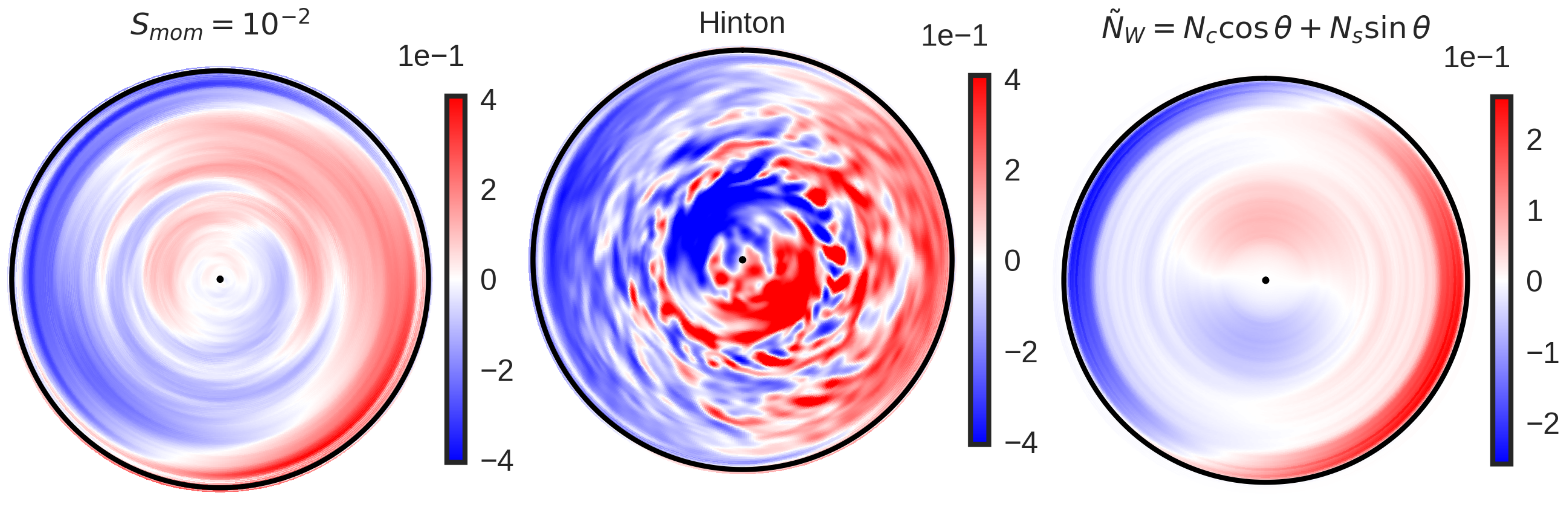}
\caption{(Left) Poloidal asymmetry of W density fluctuation $\tilde{N}_W$ obtained from the GYSELA simulation with $S_{mom}=10^{-2}$, (center) a reconstructed impurity density $\tilde{N}_W$ from the analytical expression described in Eq. (\ref{Sec2:Density_distribution}) and (right) a reconstructed impurity density $\tilde{N}_W$ from the expressions in Eqs. (\ref{Eq:Fulop}-\ref{Eq:Fulop_io}).}
\label{Fig:Asymmetry_theory}
\end{center}
\end{figure}
Figure \ref{Fig:Asymmetry_theory} represents the non-uniform distribution of W in a toroidally rotating plasma, as obtained from the GYSELA simulation on the left in Figure \ref{Fig:Asymmetry_theory}, and its comparison with analytical expressions from Eq. (\ref{Sec2:Density_distribution}) and Eqs. (\ref{Eq:Fulop}-\ref{Eq:Fulop_io}). The other parameters in the analytical expressions are evaluated from the same GYSELA simulation with $S_{mom}=10^{-2}$. Both theoretical approaches, namely the Hamiltonian in a rotating frame and the drift-kinetic equation, are in fair agreement regarding the enhanced in-out asymmetry due to the centrifugal effects.

The expression for the density distribution derived in Eq. (\ref{Sec2:Density_distribution}) shows a level of poloidal asymmetry comparable to that of GYSELA, whereas a factor of 1/2 is observed when using the expression derived from the drift-kinetic equation. The main difference between the two expressions is the inclusion of the poloidally varying electric potential $\phi(\psi, \theta)$, which arises from toroidal rotation and turbulence. The discrepancies with neoclassical predictions can be attributed to the fact that the expressions in Eqs. (\ref{Eq:Fulop}-\ref{Eq:Fulop_io}) rely on neoclassical estimates of the poloidal ions velocity \cite{Kim1991, Newton2006, Stacey2006, Wong2008}. These estimates are not always consistent with experimental observations, suggesting that other mechanisms, such as intrinsic rotation induced by turbulence, may be important in this process.

Experimental observations \cite{Bucalossi2022, Lee2022, Shengyu2022} have revealed that a strong poloidal asymmetry, which is enhanced by the centrifugal force, can lead to a significant accumulation of W in the plasma core, primarily due to the inward neoclassical convection. Given these observations, the following section will focus on exploring both neoclassical and turbulent transport of W in toroidally rotating plasma.

\section{Neoclassical and turbulent impurity transport in a toroidally rotating plasma}\label{Sec4}
\begin{figure}[H]
\begin{center}
    \includegraphics[width=0.8\textwidth]{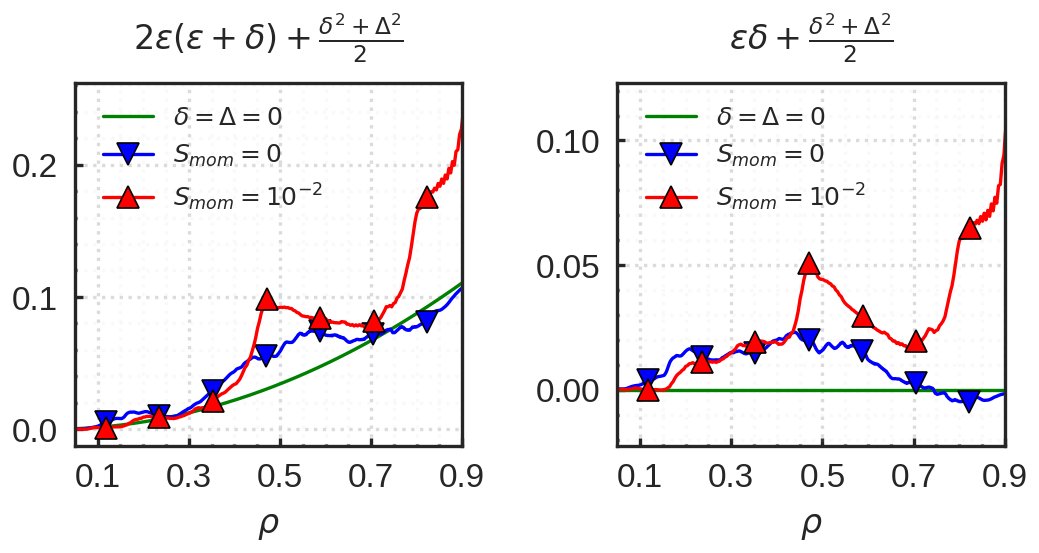}
\caption{Radial profile of the geometrical factors defined in Eqs. (\ref{Asym_geo_factor1}-\ref{Asym_geo_factor2}) obtained from the nonlinear GYSELA simulations. The green line represents the case with the uniform density distribution.}
\label{Fig:Asym_geo_factors}
\end{center}
\end{figure}

The importance of including poloidal asymmetries in neoclassical impurity transport is now well recognized, and experimentally demonstrated that a strong non-uniform distribution in the outboard region can lead to core accumulation \cite{Casson2015, Angioni2015}. An analytical expression of heavy impurity transport\textemdash Pfirsch-Schlüter transport\textemdash which takes into account such inhomogeneous poloidal density has been proposed in Ref. \cite{Angioni2014} assuming the main ions in the banana regime: 
\begin{align}
    R \langle \Gamma^{\textrm{neo}}_W \cdot \nabla \psi \rangle_\psi &\propto \Bigg[-\frac{R}{L_{ni}} -H_\textrm{neo}\frac{R}{L_{Ti}}+\frac{1}{Z}\frac{R}{L_{nW}} \Bigg]P_A, \\
    &\propto \Bigg[\Bigg(-\frac{R}{L_{ni}} -H_\textrm{sym}\frac{R}{L_{Ti}}+\frac{1}{Z}\frac{R}{L_{nW}}\Bigg) P_A -0.33P_Bf_c\frac{R}{L_{Ti}}\Bigg],
\label{Eq:neo_flux}
\end{align}
where $H_{\textrm{neo}}=H_{\textrm{sym}}+0.33(P_B/P_A)f_c$ is the temperature screening coefficient with $H_{\textrm{sym}}=-1/2$ in absence of poloidal asymmetries. The geometric factors $P_A$ and $P_B$ related to the poloidal asymmetry are defined as
\begin{align}
    P_A &= \frac{1}{2\epsilon^2} \frac{\langle B^2 \rangle}{\langle N_W \rangle}\bigg[ \bigg\langle \frac{N_W}{B^2}\bigg\rangle - \bigg\langle \frac{B^2}{N_W}\bigg\rangle^{-1}\bigg],\\
    P_B &= \frac{1}{2\epsilon^2} \frac{\langle B^2 \rangle}{\langle N_W \rangle}\bigg[\frac{\langle N_W \rangle}{\langle B^2 \rangle} - \bigg \langle \frac{B^2}{N_W}\bigg\rangle^{-1}\bigg].
\end{align}
The values of $P_A$ and $P_B$ depend on the magnitude of poloidal asymmetry, as well as its geometry, namely in-out or up-down. A simple approach to calculate these geometrical factors was proposed in the large aspect ratio limit by considering $b=B/B_0=1-\epsilon\cos\theta$ and using a similar expression to Eq. (\ref{Eq:Fulop}) for $N=1+\delta \cos \theta + \Delta \sin \theta$ with $\delta$ (in-out) and $\Delta$ (up-down) asymmetry parameters \cite{Angioni2014}. By keeping the second order accuracy, the geometrical factors can be rewritten as
\begin{align}
     \bigg\langle \frac{N_z}{B^2}\bigg\rangle - \bigg\langle \frac{B^2}{N_z}\bigg\rangle^{-1} &=2\epsilon(\epsilon+\delta) + \frac{\delta^2 + \Delta^2}{2}\label{Asym_geo_factor1}, \\
     1-\Bigg\langle \frac{B^2}{N_z} \Bigg\rangle &=\epsilon \delta + \frac{\delta^2 + \Delta^2}{2}\label{Asym_geo_factor2},
\end{align}
where we use $\langle N_Z \rangle = \langle b^2 \rangle = 1$. The radial profile of the simplified geometrical factors is illustrated in Figure \ref{Fig:Asym_geo_factors} from the nonlinear simulations. In case of a uniform density distribution corresponding to the case with $\delta=\Delta=0$, the geometrical factors are equal to $P_A=1$ and $P_B=0$ recovering the standard neoclassical transport derived in Ref. \citep{Hirshman1981}. In line with the previous results from the 2D poloidal W distribution, the toroidal rotation enhances the effect of poloidal asymmetry, especially in the outboard region $(0.7 < \rho< 0.9)$ due to the centrifugal effects. The values of $P_A$ and $P_B$ scale with $1/(2\epsilon^2)$, which are usually very large numbers. In the considered simulations, the geometrical factors in Figure \ref{Fig:Asym_geo_factors} are amplified by $1/(2\epsilon^2) \simeq 40$ at $\rho=0.5$ and $1/(2\epsilon^2) \simeq 12$ at $\rho=0.9$ yielding $P_A \simeq 4$ and $P_B \simeq 0.5$ at mid-radius. Indeed, a series of recent experimental works highlight that the poloidal asymmetry caused by external heating system, such as NBI and ion cyclotron resonance heating (ICRH) heating systems, can increase the factors $P_A$ and $P_B$ up to 100 and reduce the temperature screening coefficient $H_{\textrm{neo}}$ leading to a core accumulation near the magnetic axis \citep{Casson2020}.  

\begin{figure}[H]
\begin{center}
    \includegraphics[width=1\textwidth]{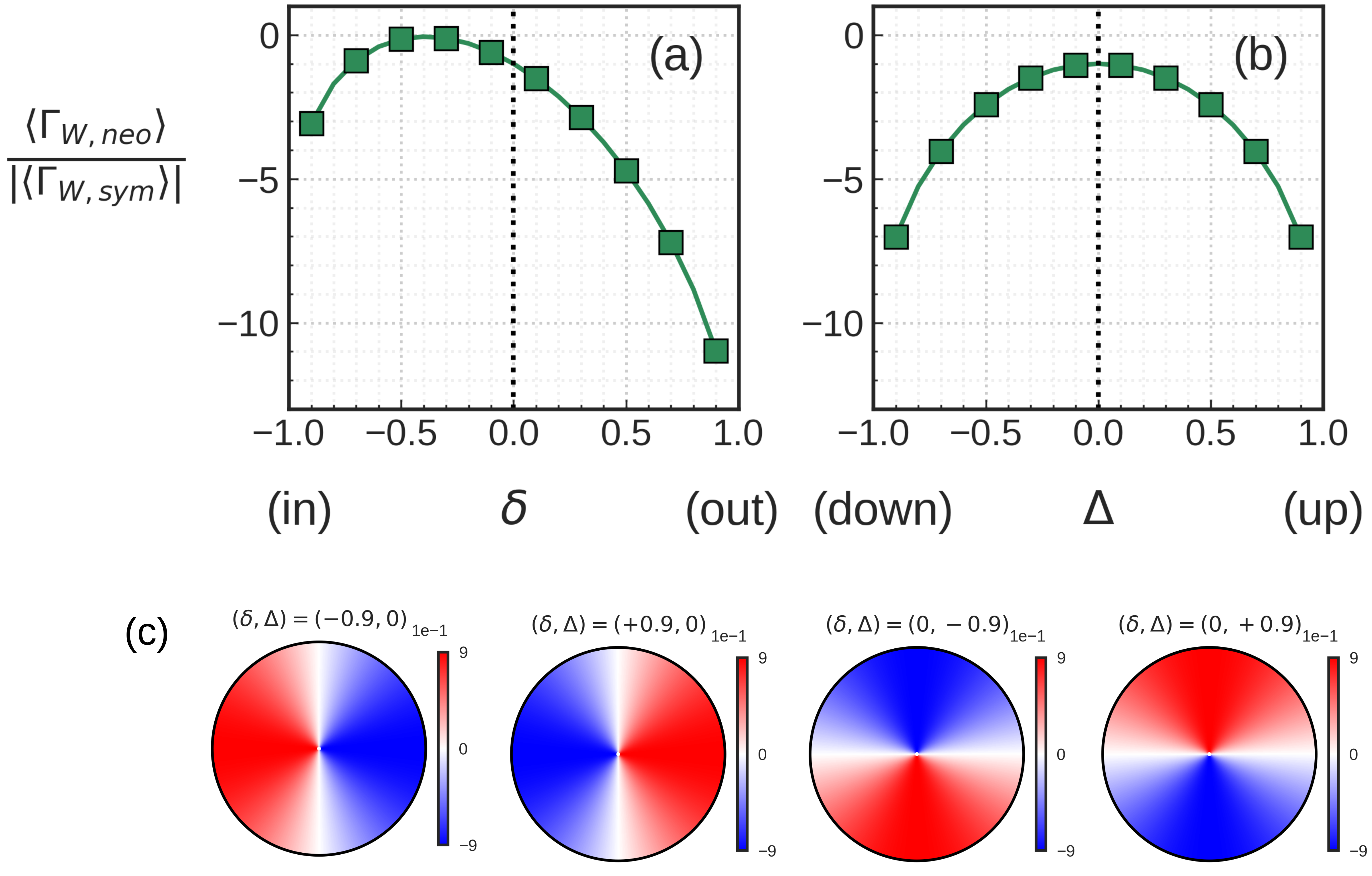}
\caption{Neoclassical W flux as a function of the $\delta$ (a) and $\Delta$ (b) asymmetry parameters. The reconstructed density (c) $\tilde{N}_W=\delta \cos\theta + \Delta \sin \theta$ is used to calculate the neoclassical flux in Eq. (\ref{Eq:neo_flux}), while other terms, i.e. $N_i, T_i, T_W, \nu_{zi}$, are taken from the GYSELA simulations. Flux is normalized to the absolute value of the symmetric case with $\delta=\Delta=0$.}	
\label{Fig:delta_Delta_neo}
\end{center}
\end{figure}

In general, neoclassical impurity transport depends on various plasma parameters for the main ions and impurities, i.e. $\Gamma_{W}^{\textrm{neo}} \propto f(N_i, N_W, T_i, T_W, \nu_{zi})$. However, unlike the standard neoclassical theory, a non-uniform distribution of impurity density has to be included to accurately predict the neoclassical flux. In Figure \ref{Fig:delta_Delta_neo}, the effect of inhomogeneous W density on neoclassical W flux is represented by performing a scan of different values in $\delta$ and $\Delta$ from $\tilde{N}_W=\delta \cos\theta + \Delta \sin \theta$. The reconstructed 2D asymmetry of W distribution is then inserted in Eq. (\ref{Eq:neo_flux}) to estimate the neoclassical flux, while other terms, measured from the GYSELA simulations, are kept constant. The poloidal asymmetry scan clearly indicates an increase of inward neoclassical flux as the asymmetry grows. In the case of up-down asymmetry (Figure \ref{Fig:delta_Delta_neo}\textcolor{blue}{b}), the trend is found to be symmetric, while the in-out asymmetry study (Figure \ref{Fig:delta_Delta_neo}\textcolor{blue}{a}) reveals a strong impact of impurity localization in the outboard region on neoclassical transport, even causing a change in the order of magnitude.

Similarly to the neoclassical particle flux defined in Eq. (\ref{Eq:neo_flux}), the turbulent W flux can be decomposed into various terms in the trace limit of impurities $(\alpha=N_W Z_W^2 \ll 1)$ \cite{Angioni2012_a} as
\begin{align}
    \frac{R\Gamma_W}{N_W} = -D_W \frac{\partial \ln{N_W}}{\partial r} + C_T\frac{\partial T_W}{\partial r} + C_u \frac{\partial v_{\varphi, W}}{\partial r} + C_p \frac{\partial q}{\partial r},
\label{Eq:turb_flux}
\end{align}
where each term represents turbulent diffusion, thermo-diffusion, roto-diffusion and curvature pinch respectively. The characteristic of these terms depends on the nature of dominant instabilities or magnetic curvature profiles that are present in the plasma. For example, the diffusion coefficient increases (decreases) with impurity charge Z for the case of Trapped-Electron-Mode (TEM) turbulence (ITG turbulence). In addition, thermo-diffusion and roto-diffusion are directed inward (outward) for TEM turbulence (ITG turbulence), while curvature-driven transport is directed inward (outward) for positive (negative) magnetic shear. The same trend was qualitatively reproduced in the recent studies with the bounced-averaged gyrokinetic code TERESA using the quasi-linear approach for the impurity transport \citep{Gravier2019, Lim2020}. A more comprehensive review on the physical mechanisms underlying each term can be found in Ref. \cite{Angioni2021}.

In the present study, the impact of toroidal rotation on turbulent flux is primarily mediated through the roto-diffusion term. The roto-diffusion term, proportional to the radial gradient of toroidal velocity, is known to be significant for W as the magnitude of the coefficient scales as $C_u \propto A_W/Z_W$.

For the analysis of the nonlinear GYSELA simulations, we adopt the following definitions for turbulent and neoclassical particle flux of W \cite{Esteve2018, Donnel2019_c},
\begin{equation}
<\Gamma_{W}^{\text{turb}}\cdot \nabla \psi >_{\psi} = \Big \langle \int d^3v \bar{F}_W\overbar{v}_E^{n\neq0} \cdot \nabla \psi\Big \rangle_\psi,
\label{Eq:GYS_turb}
\end{equation}
\begin{equation}
<\Gamma_{W}^{\text{neo}}\cdot \nabla \psi >_{\psi} = \Big \langle \int d^3v \bar{F}_W(v_{D,W} + \overbar{v}_E^{n=0}) \cdot \nabla \psi \Big \rangle_\psi,
\label{Eq:GYS_neo}
\end{equation}
where $n$ is the toroidal mode number, $\bar{v}_E$ is the $E \times B$ drift of the guiding-center and $v_{D}$ is the diamagnetic drift. Note that the axisymmetric component of the $E\times B$ drift, namely $\bar v_E^{n=0}$, also contains turbulence-driven terms, sometimes called convective cell \cite{Donnel2019_b}. In that respect, the splitting between turbulent and neoclassical contributions contained in these definitions Eqs. (\ref{Eq:GYS_turb}-\ref{Eq:GYS_neo}) should not be understood too strictly.

\begin{figure}[H]
\begin{center}
    \includegraphics[width=1.\textwidth]{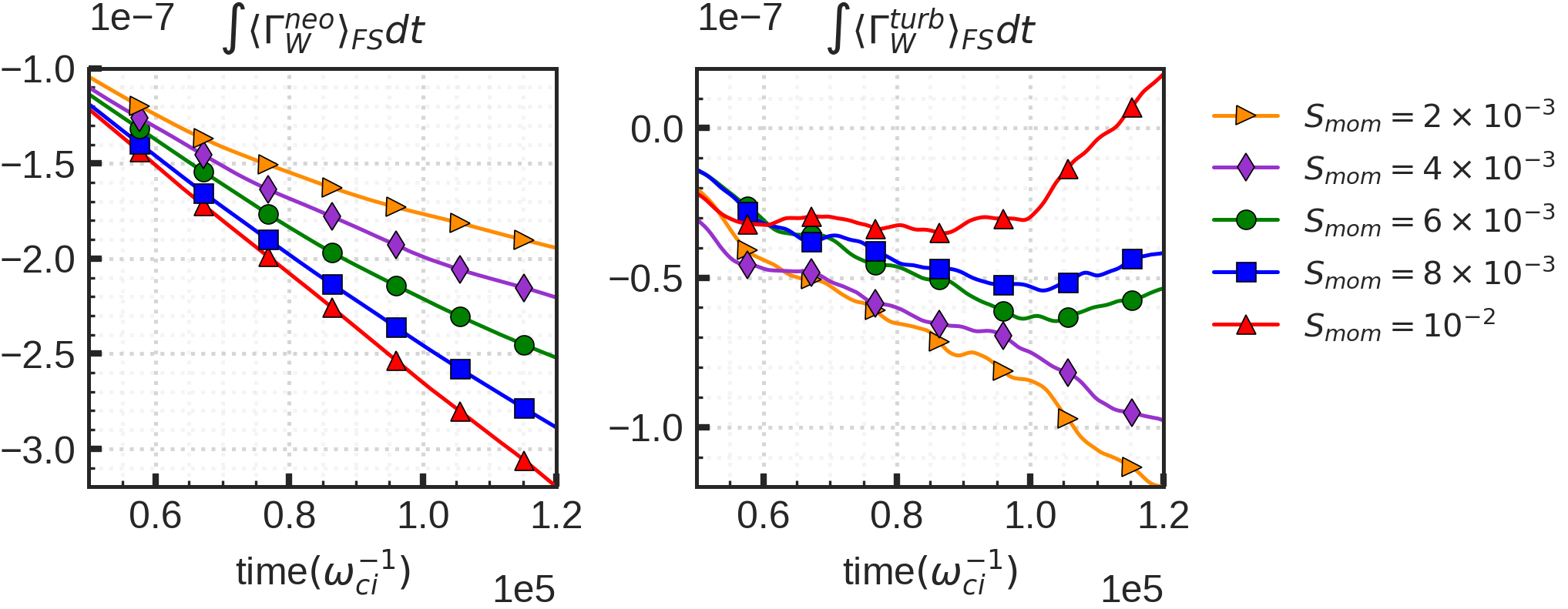}
\caption{Time accumulative neoclassical (left) and turbulent (right) W flux averaged over the radial position between $\rho=0.1$ and $\rho=0.9$ with different magnitudes of the momentum source.}
\label{Fig:Flux_accumulated}
\end{center}
\end{figure}

Figure \ref{Fig:Flux_accumulated} illustrates the cumulative turbulent and neoclassical fluxes obtained from GYSELA simulations at various levels of momentum injection. The fluxes are integrated both in time and volume within the range of $0.1<\rho<0.9$. Consistently with the previous findings in Figure \ref{Fig:delta_Delta_neo}, the neoclassical inward convection resulting from an enhanced in-out asymmetry of W density is observed to be proportional to the magnitude of toroidal rotation. A recent analysis of W accumulation in KSTAR experiments using the drift-kinetic code NEO \cite{Belli2012} revealed that neoclassical inward convection is maximized at a specific toroidal rotation value, beyond which the outward convection increases \cite{Lee2022}. The non-monotonic relationship between the toroidal rotation and neoclassical inward convection can be attributed to the different dependencies of the convection terms on the main ion density and temperature profiles, as well as collisionality. An analytical investigation of neoclassical transport shows that the temperature screening of impurities is enhanced at sufficiently high Mach number and low collisionality, while this effect decreases as collisionality increases \cite{Fajardo2023}. In the present study, however, the impact of toroidal rotation on the main profiles is found to be negligible, as depicted in Figure \ref{Fig:Radial_NTP_profiles}, and only a high collisional case with W is considered. This explains the linear relationship observed between neoclassical inward convection and the magnitude of toroidal rotation in Figure \ref{Fig:Flux_accumulated}.

Although the toroidal rotation has a direct impact on neoclassical transport owing to the presence of strong density inhomogeneity, its influence on turbulent transport in Figure \ref{Fig:Flux_accumulated} (right) is observed to be relatively small. The enhancement of outward turbulent flux resulting from the increase in toroidal rotation, as depicted in Figure \ref{Fig:Flux_accumulated}, can be attributed to the roto-diffusion and thermo-diffusion terms, as defined in Eq. (\ref{Eq:turb_flux}). Although both terms are known to drive outward convection in the case of ITG turbulence, the roto-diffusion predominates over the thermo-diffusion term, as the former is proportional to $A_W/Z_W =4.6$, while the latter is proportional to $1/Z_W = 0.025$. In fact, for heavy impurities with strong toroidal rotation, the magnitude of roto-diffusion can be as significant as the diffusive part \cite{Camenen2009}.

\begin{figure}[H]
\begin{center}
\includegraphics[width=\textwidth]{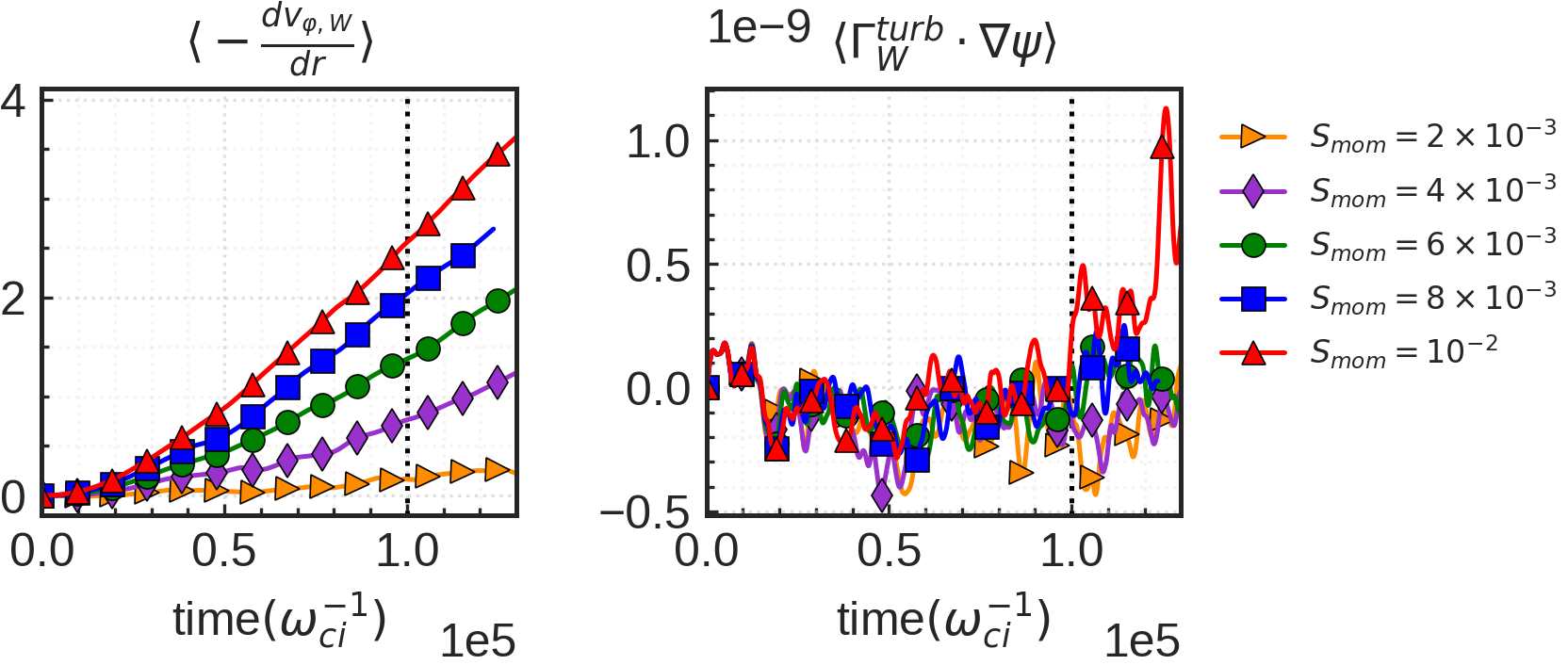}
\caption{Time evolution of the radial gradient of the toroidal velocity (left) and the time evolution of the turbulent flux of W (right) for different values of the injected momentum. The dashed vertical line indicates the time at which the effect of roto-diffusion becomes significant. }
\label{Fig:Turb_flux}
\end{center}
\end{figure}

The impact of the roto-diffusion term on the W turbulent flux becomes apparent when we compare the time evolution of the radial gradient of toroidal velocity and the turbulent flux for W, as depicted in Figure \ref{Fig:Turb_flux}. The radial gradient of toroidal rotation gradually increases over time in proportion to the injected momentum. For W, the values of $\lvert dv_{\varphi,W}/dt \rvert$ are observed to exceed unity for amplitudes above $S_{mom} = 4 \times 10^{-3}$ due to its heavy mass, whereas the gradient remains low at $\lvert dv_{\varphi,D}/dr\rvert \simeq 0.4$ for the main ions. As shown in Figure \ref{Fig:Flux_accumulated}, the outward turbulent flux emerges at $t=10^{5}\omega_{ci}^{-1}$ for amplitudes greater than $S_{mom}=6 \times 10^{-3}$. The results demonstrate that the influence of the roto-diffusion term becomes more significant compared to other terms in Eq. (\ref{Eq:turb_flux}) as the absolute value of $\lvert dv_{\varphi,W}/dr \rvert$ gradually increases. Additionally, it is noteworthy that centrifugal effects also have a non-negligible impact on other convection terms, particularly in the case of heavy impurities \cite{Angioni2012_a}. Consequently, the sign and magnitude of the turbulent flux depend on the interplay between different convection terms, which typically results in an outward flux in the case of ITG turbulence.
\begin{figure}[H]
\begin{center}
    \includegraphics[width=0.4\textwidth]{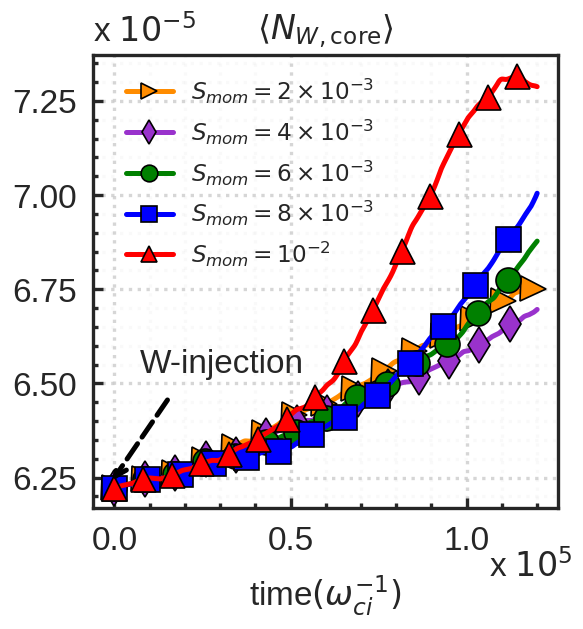}
\caption{Core W-accumulation in time for different amplitudes of the toroidal momentum source. The core density $N_{\textrm{W, core}}$ is integrated between $\rho=0.1$ and $\rho=0.3$. The strong toroidal velocity leads to an enhanced core accumulation.}	
\label{Fig:Core_accumulation}
\end{center}
\end{figure}

Figure \ref{Fig:Core_accumulation} presents the W accumulation obtained from the total flux, $\Gamma_W^{tot}=\Gamma_W^{neo} + \Gamma_W^{turb}$, within the radial range of $\rho=0.1$ and $\rho=0.3$ for different values of the injected momentum. No significant changes in the core W density are observed up to the time $t=0.5 \times 10^{5}\omega_{ci}^{-1}$, which corresponds to the momentum transfer time from the main ions to W. Following this transfer time, a more significant accumulation of W is observed, approximately $20\%$ of increase in the $S_{mom}=10^{-2}$ scenario. The central accumulation of W in GYSELA simulations are commonly observed even in the absence of external toroidal rotation \cite{Lim2021}. The numerical results obtained in the present study are consistent with experimental evidence from different tokamaks \cite{Casson2020, Angioni2015, Lee2022} where central NBI heating yields a central deposition of W. To prevent the accumulation of W within the core region, the proper use of ICRH and ECRH systems can be helpful in controlling the W within the core region \cite{Dux2003, Puiatti2006, Valisa2011, Sertoli2017, Goniche2017, Angioni2017, Bilato2014, Shengyu2022}. This is achieved by flattening the density profile of main ions and enhancing the temperature gradient of the bulk plasma, which facilitates the outward turbulent diffusion of W and an increase of the temperature screening effect. In addition, the temperature anisotropy that arises from ICRH is known to reduce poloidal asymmetry, thereby decreasing inward neoclassical convection \cite{Bilato2014, Bilato2017}. 
\section{Conclusion}\label{Sec5}
In the present paper, the effects of toroidal rotation on both turbulent and neoclassical W transport are addressed using the nonlinear, global, flux-driven 5D gyrokinetic code GYSELA. The injection of toroidal momentum via the external source term drives W to the supersonic regime, while the toroidal velocity of the main ions remains low. By leveraging the multi-species collisional operator and external source terms implemented in GYSELA, both turbulent and neoclassical W transport in a toroidally rotating plasma are treated self-consistently, thereby capturing the interplay between these two channels of W transport.

Through a series of nonlinear simulations involving different levels of momentum injection, it has been observed that centrifugal forces cause W to accumulate in the outboard region, creating an in-out poloidal asymmetry. This asymmetry has been found to significantly enhance the neoclassical inward convection, leading to a central accumulation of W in the case of strong plasma rotation. The core W accumulation accelerated by the centrifugal force and the resulting poloidal asymmetry, cannot be offset by the turbulent flux. However, as the momentum injection continues, the roto-diffusion term becomes as significant as the diffusive part once the radial gradient of the rotation velocity exceeds a certain threshold, driving outward turbulent flux in the case of ITG turbulence.

The numerical results obtained from the present study are in qualitative agreement with both analytical and experimental predictions. However, for a more comprehensive modelling of W transport in GYSELA, it is necessary to include the time evolution of the main ions density and temperature profiles in response to other heating systems, such as ICRH and ECRH, which are known to mitigate the core W accumulation by reducing poloidal W asymmetry and increasing temperature screening effect.

\begin{acknowledgments}
This work has been carried out within the framework of the EUROfusion Consortium, partially funded by the European Union via the Euratom Research and Training Programme (Grant Agreement No 101052200 — EUROfusion). The Swiss contribution to this work has been funded by the Swiss State Secretariat for Education, Research and Innovation (SERI). Views and opinions expressed are however those of the author(s) only and do not necessarily reflect those of the European Union, the European Commission or SERI. Neither the European Union nor the European Commission nor SERI can be held responsible for them. The simulations presented herein were carried out on the CINECA Marconi supercomputer under FUA35 GSNTITE project.
\end{acknowledgments}
\appendix
\numberwithin{equation}{section}
\medskip
\bibliographystyle{unsrt}
\bibliography{9_bib}

\end{document}